\documentclass{vgtc}                     

\usepackage{times}                     
\usepackage{mathptmx} 

\onlineid{0}

\vgtccategory{Research}

\vgtcpapertype{evaluation}

\title{Understanding Expert Exploration in EHR Visualization Tools:\texorpdfstring{\\}{} The ParcoursVis Use Case}

\author{Assor Ambre\thanks{e-mail: first-name.last-name@inria.fr. }\\ %
        \scriptsize Inria %
\and Jean-Daniel Fekete$^*$\\ 
    \scriptsize Inria} %

\abstract{%
\am{We introduce our ongoing work toward an insight-based evaluation methodology aimed at understanding practitioners' mental models when exploring medical data.
It is based on ParcoursVis, a Progressive Visual Analytics system designed to visualize event sequences derived from Electronic Health Records at scale (millions of patients, billions of events), developed in collaboration with the Emergency Departments of 16 Parisian hospitals and with the French Social Security.}
Building on prior usability validation, our current evaluation focuses on the insights generated by expert users and aims to better understand the exploration strategies they employ when engaging with exploration visualization tools. We describe our system and outline our evaluation protocol, analysis strategy, and preliminary findings. Building on this approach and our pilot results, we contribute a design protocol for conducting insight-based studies under real-world constraints, including the availability of health practitioners whom we were fortunate to interview. Our findings highlight a loop, where the use of the system helps refine data variables identification and the system itself. We aim to shed light on generated insights, to highlight the utility of exploratory tools in health data analysis contexts.
}


\usepackage{tabu}                      
\usepackage{booktabs}                  
\usepackage{enumitem}
\usepackage{tikz}
\usetikzlibrary{graphs, calc}
\usetikzlibrary{arrows.meta, positioning, fit, shapes.multipart}
 
\usepackage{algorithmic}
\usepackage{algorithm}
\usepackage{array}
\usepackage[caption=false,font=normalsize,labelfont=sf,textfont=sf]{subfig}
\usepackage{cite}
\usepackage[svgnames,table]{xcolor}
\usepackage{xspace}
\usepackage[
    pagebackref,
    bookmarks,
    colorlinks=true,
    linkcolor={black},
    citecolor={black},
    urlcolor={black}]{hyperref}
\usepackage{comment}
\usepackage{siunitx}
\usepackage{flushend}

\newcommand{\jdf}[1]{\textcolor{black}{#1}}

\newcommand{\am}[1]{\textcolor{black}{#1}}
\newcommand{\IT}{Icicle tree\xspace}

\begin{document}
\maketitle
\section{Introduction}
Healthcare professionals, including physicians, are typically trained to study predefined patient cohorts to answer targeted clinical or research questions using traditional confirmatory statistical analyses. In contrast, exploratory data analysis (EDA) remains relatively unfamiliar to many of them~\cite{kim2024phenoflow}. This gap in experience, combined with the complexity of novel visualization interfaces, can affect how practitioners engage with tools designed to support open-ended exploration. In this work, we investigate whether and how domain experts can leverage visualization tools for exploratory analysis of healthcare data. Specifically, we aim to understand the cognitive and analytical strategies healthcare practitioners adopt when interacting with exploratory systems to perform open-ended tasks. We also want to document the benefits of visualization tools to practitioners since they are seldom reported~\cite{Fekete2008}.

We developed ParcoursVis~\cite{ParcoursVis}, a progressive visual analytics (PVA)~\cite{PDABook} system for exploring large-scale sequences of care events, in collaboration with the Parisian Emergency Departments (ED) and the French National Health Insurance. ParcoursVis enables interactive navigation through the care pathways of millions of patients. The system builds on prior work, such as EventFlow~\cite{Wongsuphasawat:2012:TVCG}, to extend its validated user interface and visualizations, but is optimized for scalability~\cite{ParcoursVis}, offering a comprehensive exploration of care pathways.

Evaluating the effectiveness of such a tool in real-world expert settings remains a significant challenge~\cite{Wang:2022:CGF}. While conventional usability studies can assess interface efficiency, they often fail to capture the depth of analytical reasoning or knowledge generation~\cite{plaisant2007promoting}. Longitudinal studies, though better suited to capturing real-world value, are often difficult to lead in a pressured expert's domain like emergency care~\cite{Wang:2022:CGF, plaisant2007promoting}. As a compromise, we build on our previous usability results and adopt an insight-based evaluation methodology. Our approach focuses both on the exploratory strategies and mental models practitioners use when interacting with complex care data and on the insights generated, in terms of relevance, actionability, and clinical depth.  The latter is especially important, as insights are rarely disclosed and discussed in similar studies. We argue that, in applied settings like healthcare, highlighting the relevance and actionability of these insights is essential to demonstrate how exploratory visualization tools can improve real-world processes.

\am{While the system’s technical design enables fast and large-scale exploration, this paper does not contribute novel visualization or interaction designs. Instead, our contribution lies in the methodology for evaluating how such tools support expert sensemaking, exploratory strategies, and actionable insight generation in real-world settings and for millions of patients. Our evaluation protocol consists of two main sessions. In Session 1, participants complete guided usability tasks before engaging in open-ended exploration to generate and discuss preliminary insights. In Session 2, these insights are reviewed by independent clinical experts and the participants themselves to assess their validity, relevance, and actionability, enabling cross-participant comparison and validation.}

This paper briefly describes the ParcoursVis system and contributes: (1) the design of an insight-based study and our analytical strategy for characterizing insights and experts' exploration strategies; and (2) lessons learned from initial pilot interviews with the head of an ED and an emergency doctor, along with envisioned refinements to the study protocol. By sharing our approach and lessons learned, we aim to inform ongoing conversations in the health visualization communities about the evaluation of exploration tools in expert domains and provide an initial protocol.

\section{Background}

Our work relates to temporal event sequences visualization
and user evaluation of health visualization systems. 

\subsection{Temporal Event Sequences}

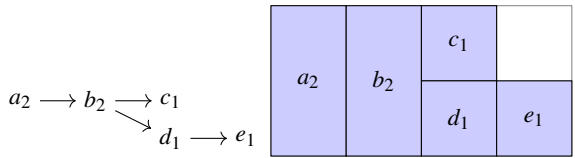
\begin{figure}[tb]
    \centering
    \begin{tikzpicture}
        \graph [branch down=5mm, math nodes] { a_2 -> b_2 -> { c_1, d_1 -> e_1} };
    \end{tikzpicture}
    \begin{tikzpicture}[fill=blue!20]
        \draw[help lines] (0, 0) grid (4,2);
        \path (0.5,1) node(a) [rectangle,minimum height=2cm,minimum width=1cm,draw,fill]    {$a_2$}
              (1.5,1) node(b) [rectangle,minimum height=2cm,minimum width=1cm,draw,fill]    {$b_2$}
              (2.5,1.5) node(c) [rectangle,minimum height=1cm,minimum width=1cm,draw,fill]  {$c_1$}
              (2.5,0.5) node(d) [rectangle,minimum height=1cm,minimum width=1cm,draw,fill]  {$d_1$}
              (3.5,0.5) node(e) [rectangle,minimum height=1cm,minimum width=1cm,draw,fill]  {$e_1$};
    \end{tikzpicture}

    \caption{ESS Aggregation of EHR Sequences as implemented by EventFlow and ParcoursVis. Patient \textit{A} took the treatment sequence $\{a, b, c\}$ in that order, and patient \textit{B} took $\{a, b, d, e\}$. The left side shows the \emph{prefix tree} of these sequences, and the right side shows the \IT visualization where node heights encode frequency.}
    \label{fig:ESS}
\end{figure}

Like EventFlow~\cite{Monroe:2013:TVCG}, ParcoursVis visualizes patient care pathways using aggregation and filtering of event sequences. 
According to the taxonomy in Wang et al.’s survey of Electronic Health Records (EHRs) visualization~\cite{Wang:2022:CGF}, these works fall into the ``Event Sequence Simplification'' (ESS) family of techniques (as described in~\autoref{fig:ESS}), i.e., any technique used for reducing the visual complexity of event sequences in aggregated display overviews. 
This family of techniques is the only one focused on visualizing an overview of the event sequences. Others rather focus on one patient, a set of patients~\cite{Wang:2022:CGF}, spatial areas sharing some characteristics, or on sequence mining~\cite{liu2016patterns, perer2014frequence, DBLP:journals/tvcg/VrotsouN19, Stolper:2014:TVCG} (i.e., automatically discovering frequent and important temporal patterns).

\begin{figure}\centering
    \includegraphics[width=0.95\linewidth]{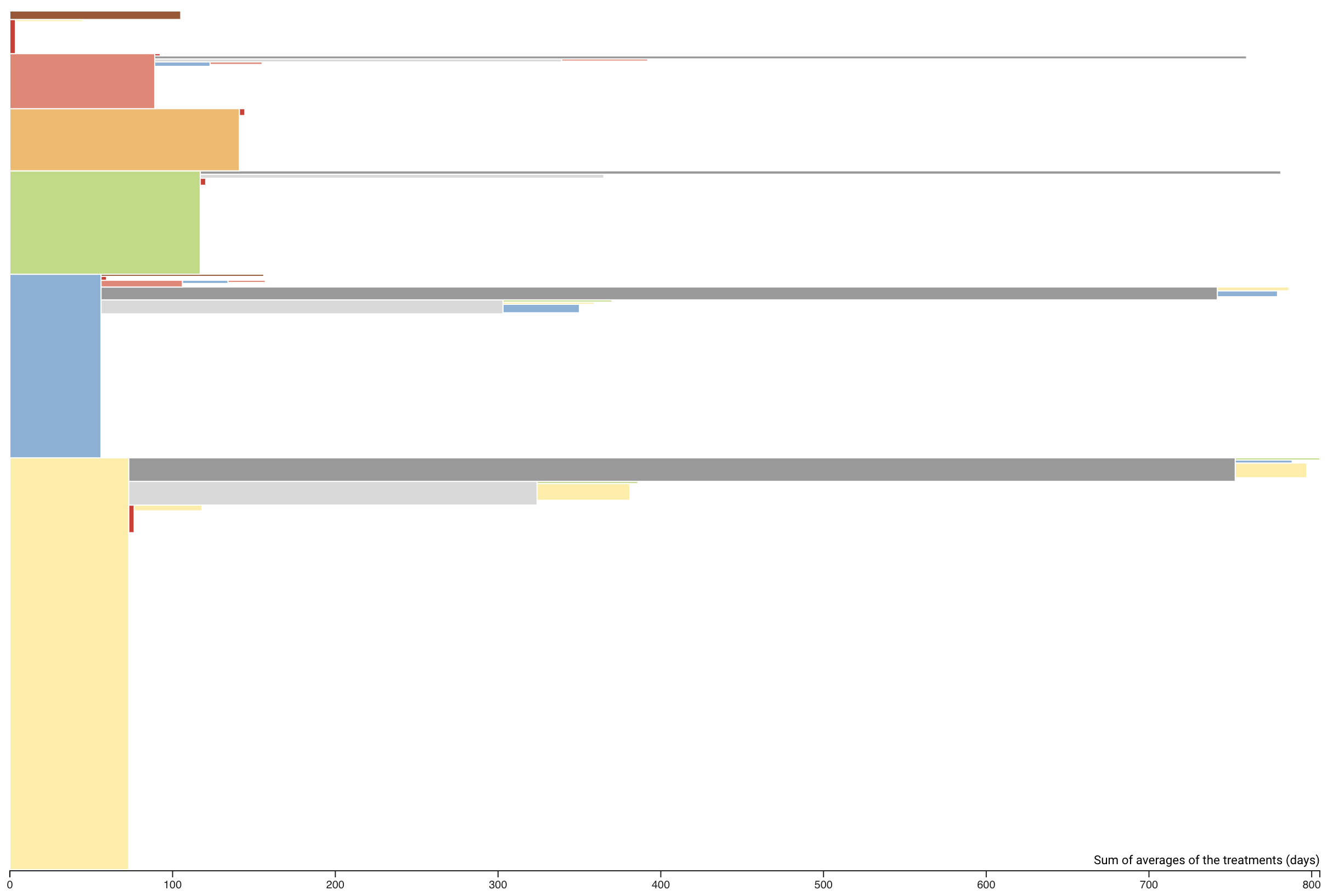}
    \caption{An aggregated \IT displaying 10M treatment sequences. 
    Node height indicates how frequently each event occurs. 
    Node children are on the right of their parent, sorted by height. 
    Node width reflects the estimated duration of the event.
    Color encodes the event type. 
    The first seven nodes on the left represent initial treatment choices.
    A treatment pathway is trackable by moving through nodes from left to right.}\label{fig:parcoursvis}
\end{figure}

ParcoursVis reuses the \am{core visual design } from EventFlow (\autoref{fig:parcoursvis}), which has been validated by several use cases for visualizing medical temporal event sequences~\cite{liu2016patterns, perer2014frequence, Stolper:2014:TVCG, Fisher:2012:CGA}.
\jdf{This design relies on EventFlow's visual encoding, aggregating a large number of event sequences into a \emph{prefix-tree} visualized as an \IT. We focus on understanding how existing, validated designs support knowledge discovery at the scale of real medical datasets.}

\am{While EventFlow performs well for small datasets ($< 20$k patients), it cannot scale to millions of patients.} \jdf{Random sampling can mitigate the scalability issue, but it could also} hide rare but crucial sequences that are important to practitioners. To handle larger datasets without losing rare sequences, ParcoursVis uses Progressive Visual Analytics (PVA)~\cite{PDABook, Stolper:2014:TVCG}; it incrementally builds the aggregated prefix tree, offering intermediates, improving visual feedback during the computation \jdf{every second or so until the whole dataset is processed. With PVA, ParcoursVis can scale to millions of patients \jdf{and be applied to datasets 3-5 orders of magnitude larger than the other EHR visualization tools.} 
Yet, our main contribution here lies in the methodology used to evaluate how such a tool supports expert sensemaking and insight generation at real scales.}

\subsection{Mental Models and Evaluation Strategies for EHR\texorpdfstring{\\}{} Visualization Tools}
\label{sec:studies}

Evaluating the effectiveness of visualization tools for EHRs presents unique challenges due to the complexity of the data and interfaces, as well as the specialized needs of domain experts. Many existing systems are evaluated through domain expert feedback~\cite{Wang:2022:CGF}, which has become increasingly integrated across the development lifecycle, particularly since 2018. Approaches such as grounded evaluation, which emphasize close collaboration between designers and health professionals during prototyping, refinement, and late-stage testing, are emerging as standard practice in the field. Other widely adopted evaluation methods include longitudinal designs, such as the Multi-Dimensional In-Depth Long-Term Case Studies (MILCS) methodology~\cite{MILCS}. \am{MILCs combine multiple qualitative and quantitative approaches (e.g., observations, interviews, interaction logging) to study a small set of expert users over extended periods in their real work environments. The goal is not only to refine the tool, but also to assess the extent to which it contributes to the users’ own professional success.} While MILCS has demonstrated methodological robustness and has been successfully applied in various contexts, it poses significant practical challenges in real-world settings. One of them is the substantial time commitment required from physicians, which is something that is often difficult to secure given their clinical responsibilities. Furthermore, the methodology relies on participants’ autonomy and their ability to access shared data environments. In our case, we use a secure data warehouse with access to specific datasets that have been approved by the relevant authorities. However, not all physicians have access to this development environment, and these operational constraints limit the autonomous access of our target evaluation users.

Also, while these approaches bring valuable contextual insights, they often focus on usability rather than on how tools support knowledge generation. This has led to growing interest in insight-based evaluation approaches, which are particularly useful for assessing tools that support discovery rather than task completion. Studies by Saraiya et al.~\cite{saraiya2005insight} propose protocols involving think-aloud techniques, video analysis, and expert coding to quantify insight complexity, domain relevance, and hypothesis development. 
These approaches have proven valuable for evaluating tools intended for expert-driven exploration of real-world data, like those found in EHR settings, where both confirmatory and unexpected insights matter. 
\am{In our context, we adopt Saraiya et al.’s definition of an insight as \emph{an individual observation about the data by the participant, a unit of discovery} made during exploration, concerning patterns, anomalies, or relationships in the data that the participant considers relevant or noteworthy in their domain. For example, an insight might be the observation of a treatment sequence that deviates from standard guidelines or detecting a higher-than-expected frequency of transfers between two specific care zones.}

However, while insight-based evaluations capture some of the outcomes of exploration (i.e., the insights users might generate), they often focus on surface-level metrics such as the number or correctness of insights~\cite{saraiya2005insight}. As a result, the insights themselves are rarely disclosed, discussed, or examined in terms of their potential use in real-world settings. Furthermore, such evaluations provide limited visibility into the cognitive processes that produce these insights. In this context, researchers have drawn on sensemaking models like Pirolli and Card’s loop~\cite{pirolli2005sensemaking} and Klein et al.’s Data-Frame Theory~\cite{klein2007data}, highlighting the interplay between bottom-up (data-driven) and top-down (frame-driven) exploration. For instance, studies such as~\cite{choi2019visual} show how juxtaposing actual vs.\ expected outcomes can help refine frames, a design-oriented approach that emphasizes how specific features can guide sensemaking. We adopt a broader perspective and propose a dual approach: first, we use an insight-based evaluation to assess the quality and actionability of produced insights; second, we aim to capture the cognitive strategies underneath insight generation throughout the entire exploratory session (i.e., how users form, refine, and shift mental models). This will allow us to link the cognitive processes to insights' characteristics such as value, actionability, or their confirmatory or unexpected nature.

\am{Rather than proposing new visualization techniques or system features, our goal is to advance methodologies for evaluating how visualization tools support complex exploratory reasoning in real-world clinical contexts.}

\section{ParcoursVis}

To ensure the system can be evaluated in realistic conditions, we designed ParcoursVis~\cite{ParcoursVis} to operate on large-scale health data. This requires addressing both performance and scalability challenges. Our benchmark of ParcoursVis using a synthetic Prostate Adenoma treatment dataset with hundreds of events per patient shows that, on a computer using six cores in parallel, it processes up to 13 million patient sequences per second while remaining interactive. This performance is 3--5 orders of magnitude higher than similar systems, depending on the measure. 
In this section, we summarize the implementation, design, use cases, and tasks handled by our system.

\subsection{Implementation}
ParcoursVis combines JavaScript for the front end and C++ for the back-end processing. 
The front end, consisting of approximately 3000 lines of JavaScript, uses D3.js and Vue.js for dynamic visualizations in a web browser. 
The back end, written in C++ ($\approx 3000$ lines), is used as a Python module and performs the computation required for aggregating and processing data.
ParcoursVis applies a combination of data structure optimizations, parallelization, and progressive computation to handle large datasets with billions of events while remaining interactive.

\subsection{Use Cases}\label{sec:usecases}
ParcoursVis is currently used in two use cases that we describe below. We are also working on more use cases using the same code base to explore specific cancer treatments at the scale of France.

\noindent\textbf{Emergency Departments (ED):}  
ParcoursVis is used to explore patient care pathways across three public Emergency Departments (ED) in Paris ($\approx 1.5$M patients). With increasing patient volumes over the last 15 years, there is a need to identify critical bottlenecks in emergency care. 
The ED staff uses \am{ORBIS}, a centralized system to log patient records, including demographics, procedures, and care locations. Data is saved every 15 minutes, enabling near-real-time granularity. The system processes event sequences where each low-level event corresponds to ED locations: Waiting Area (non-medical holding space), Triage Room (initial assessment), Standard Cubicle (general consultations), Technical Cubicle (specialized exams), Shock Room (intensive care unit), and UHCD (short-term hospitalization). Additional attributes include age, hospital, and triage score (1 being critical and 5 non-urgent).
\jdf{We call \emph{low-level} the events that are directly stored in the database and not synthesized or summarized.}
\am{For this use case \emph{ParcoursVis} is deployed as a widget within the hospital’s JupyterLab environment instead of as a Web application for security reasons, as discussed in \autoref{sec:studies}.}

\noindent\textbf{Non-cancerous Prostate Adenoma Treatments:}  
ParcoursVis is applied to analyze non-cancerous prostate adenoma treatments using data from the French Social Security’s reimbursement database (2.5M patients). This data includes drug purchases (e.g., alpha-blockers) and surgeries. A version of ParcoursVis showing pathways for 10M synthetic patients treated for non-cancerous prostate adenoma is available at \href{https://parcoursvis.lisn.upsaclay.fr/}{parcoursvis.lisn.upsaclay.fr}. The system is open source, available at \href{https://gitlab.inria.fr/aviz/parcoursvis}{gitlab.inria.fr/aviz/parcoursvis}.

\smallskip

In ParcoursVis, low-level events\jdf{, provided by a database} (e.g., drug purchases or patient locations in the ED), are aggregated into higher-level events to simplify analysis. For example:

\begin{itemize}[nosep]
    \item In the ED use case, multiple visits to the same location within a short period are combined into one high-level event with a summarized duration.
    \item In the prostate treatment use case, multiple consecutive purchases of the same medication are grouped into a single event (e.g., alpha-blockers for a certain duration), clarifying treatment pathways.
\end{itemize}

These aggregations help users visualize complex sequences and identify key patterns without getting lost in the details.

\subsection{Views}

\textbf{Main View:} Displays the aggregated tree of patients’ care pathways, as explained in \autoref{fig:parcoursvis}. 
The “Prefix View” focuses on sequences leading to or following a specified event.

\noindent\textbf{Control Panel:} Provides access to the “Overview,” “Detail,” “Filtering,” and “History” tabs. The “Overview” shows event hierarchies, and the “Detail” tab visualizes patient data distributions like age and duration of stay or drug purchase. The “Filtering” tab allows selection of patient subsets, while the “History” tab tracks and compares previous views.

\subsection{Tasks Supported}
\label{sec:tasks}
\noindent\textbf{Simple Tasks}
\begin{itemize}[nosep]
    \item Overview: Visualize all temporal event sequences. 
    \item Zoom and Filter Events: Visualize sequences with a given prefix or filter event types.
    \item Details on Demand: Visualize attributes and distributions related to a specific sequence (e.g., age or duration).
\end{itemize}

\noindent\textbf{Advanced Tasks:}
    \begin{itemize}[nosep]
    \item Filter by Attribute: Filter data based on attributes such as sequence length or patient age.

    \item Configure View: Align the visualization to a specific sequence of events.

    \item Abstracting: Regroup multiple events into a super type using a hierarchy.

    \item History and Comparison: Compare different tree states and navigate through past views.
\end{itemize}

\noindent\textbf{Analytical Tasks:}
\begin{itemize}[nosep]
    \item Application Questions: Identify trends, frequent, or critical pathways, and find actionable strategies.

    \item Data Questions: Detect data errors and suggest data management methods.
\end{itemize}

\subsection{First Formative Usability Results}

We ran a first formative study to assess the usability of ParcoursVis.
We interviewed five expert healthcare professionals: a referring doctor, two emergency physicians, two clinical research data analysts, as well as a researcher in computer science interested in simulating patient flow in EDs. 
We asked them to complete some of the tasks listed in~\autoref{sec:tasks} using a think-aloud approach, followed by a semi-structured interview, with all sessions audio-recorded.

Our participants were able to complete the tasks without major difficulty using ParcoursVis, \am{using as a baseline the ability to finish each usability task unaided and without excessive delays.} 
These results indicate that the tool is usable in its current version and suitable for supporting exploratory analysis by domain experts.
They also reported some usability issues and made suggestions that we addressed (e.g., improving slider labels and legends). Despite these initial issues, all participants found the tool engaging and useful to them. None of them noticed that the visualization was progressive, using it like a regular interactive tool.

One referring doctor described spending hours interacting with the tool, likening it to a video game, due to the engaging nature of the exploration. \am{Participants surfaced unexpected insights, such as prolonged stays in critical care areas that deviated from hospital guidelines, e.g., certain types of rooms intended to have minimal occupancy and short stay durations to maintain patient flow were, in practice, occupied for much longer periods. They also identified a higher-than-expected rate of treatment resumption after surgery, sometimes involving additional procedures}. The clinical researchers also saw clear potential for using the tool to support the formulation and investigation of clinical hypotheses. This motivated us to continue with an insight-driven usability study. 

\subsection{ParcoursVis Applicability}
\am{
More broadly, these first findings stressed the potential of ParcoursVis not only as a research tool but also as a system that could be integrated into hospital dashboards or public health monitoring platforms. Based on participant feedback, we see realistic deployment in contexts where care pathway optimization, resource planning, or policy evaluation are critical.
For instance, within hospital settings, healthcare professionals could employ ParcoursVis to examine temporal trends in patient pathways across comparable periods and at multiple aggregation levels (e.g., monthly, yearly). Such analyses could support the estimation of the most probable care trajectories and their associated waiting times, facilitating communication with patients on-site. Beyond the hospital, ParcoursVis could function as a strategic decision-support tool, enabling administrators and heads of departments to investigate high-level questions related to hospital management and service organization, such as comparing operational performance across departments or optimizing resource allocation. Such analyses could also be used to assess the alignment of care pathways and resource usage with institutional policies and national healthcare guidelines.}

\jdf{Additionally, ParcoursVis has been useful at highlighting data quality issues. The health institutions we collaborate with have been responsive to fixing the issues we found. They are also considering collecting more data to better answer medical questions raised by healthcare professionals on care pathways. }

\section{Usability and Insight-Based Study}
\begin{figure*}[!t] 
\centering
\resizebox{\textwidth}{!}{%
\begin{tikzpicture}[
  node distance=8mm and 12mm,
  >=Latex,
  stage/.style={rounded corners, draw, thick, align=left, inner sep=2mm, font=\footnotesize, minimum width=42mm},
  small/.style={rounded corners, draw, thick, align=left, inner sep=2mm, font=\footnotesize, minimum width=38mm}
]

\node[stage] (pilot) {\textbf{Pilot}\\
2 ED physicians\\
Protocol refinement};

\node[stage, right=of pilot] (sessionone) {\textbf{Session 1}\\
5 to 10 ED staff\\
Onboarding + exploration\\
insights generation + self-ratings};

\node[small, above right=of sessionone] (expert) {\textbf{Session 2.1}\\
Experts' review};
\node[small, below right=of sessionone] (participants) {\textbf{Session 2.2}\\
Participants' review};

\node[stage, above right=of participants, yshift=3mm] (final) {\textbf{Validated Insight Set}\\
Triangulated ratings\\
Classified by characteristics};

\draw[->, thick] (pilot) -- (sessionone);
\draw[->, thick] (sessionone.north east) -- (expert.west);
\draw[->, thick] (sessionone.south east) -- (participants.west);
\draw[->, thick] (expert.east) -- (final.west);
\draw[->, thick] (participants.east) -- (final.west);

\end{tikzpicture}%
}
\caption{\am{Study flow from pilot to final validated insight set. Session~2 combines experts' review and participants' review to rate insights from Session~1.}}
\label{fig:flow}
\end{figure*}
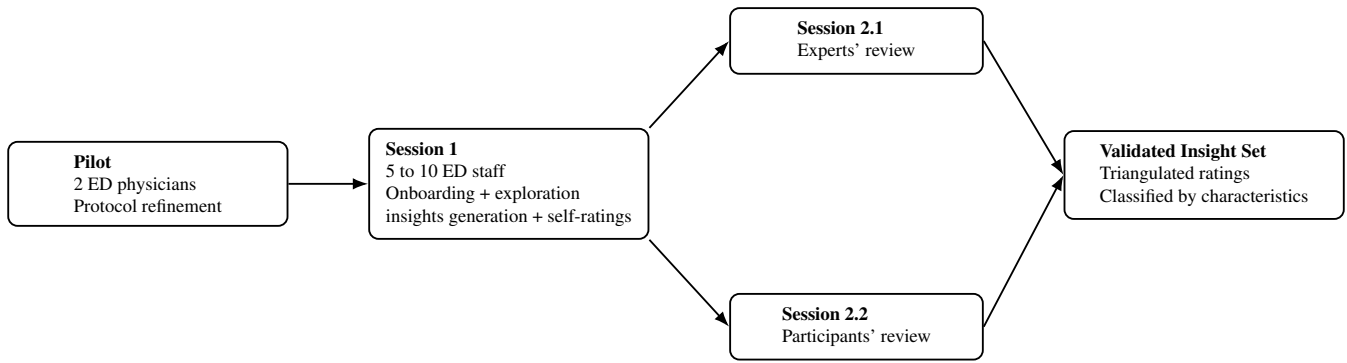

\label{sec:studyDesign}
In this section, we present the design of the second study of ParcoursVis, which aims to capture user insights and uncover the mental models of domain experts when using visualizations for exploration. We piloted this protocol and will convey lessons learned and envisioned refinements in the next section.

\subsection{Participants}
This study is designed to interview field experts, including physicians and nurses working in emergency departments in Parisian hospitals. We target between 5 and 10 participants (not including the pilots) to ensure diversity of perspectives and increase the reliability of the results. 
So far, experts have been very willing to participate in our study, despite their high workload.
Participants are audio-recorded during the whole study. The study has been approved by the experimenter \am{laboratory's IRB (i.e., ethical comittee)}.

\subsection{Session 1: In-Person Study (1h30)}

Before the session, participants receive a short video ($\approx 2$ minutes) explaining the system and study context to ensure a shared baseline of understanding.

\begin{itemize}[nosep]
    \item \textbf{Introduction and Demographics (10 min):} 
    \am{The session opens with a brief explanation of the study’s purpose and structure. We then review the core features of the tool and introduce the dataset used in the session. In particular, we present the different event types to participants in a way that is grounded in their knowledge. For instance, for the ED use case, we start from the perspective of the centralized application they use to log patient information and locations. Because the events composing the event sequences visualized correspond to the abstract zones already used in ORBIS, participants are familiar with their provenance and meaning.} Participants then complete a short demographic survey covering their professional background (e.g., role, years of experience, familiarity with data and visualization tools). To further contextualize their perspectives, participants are asked a few open-ended questions about their interest and experience with patient care pathways in emergency departments. These include questions about their professional motivation for studying patient flows, prior knowledge of care trajectories, and what they expect to gain from visualizing such data. Participants are informed that the insights they surfaced will be analyzed, presented, and discussed as part of our evaluation, with a focus on their relevance and potential clinical value.

    \item \textbf{On-boarding (20 min):} 
    Participants are introduced to the core features of the tool and complete a series of guided tasks designed to assess usability. These tasks include basic interactions such as filtering data, hiding event types, adjusting parameters, and navigating the change history. During this phase, participants are encouraged to \textit{think aloud}, i.e., to verbalize their thoughts, intentions, and impressions as they engage with the tool. This think-aloud protocol helps reveal their understanding, expectations, and any usability issues as they arise in real-time.
   This session serves both to onboard participants, giving them hands-on familiarity with the tool, and to capture any remaining usability concerns.

    \item \textbf{Exploratory Insight Generation (40 min):}  
    Participants are asked to freely explore the dataset and \textit{think aloud} as they examine real patient care trajectories using ParcoursVis. They are encouraged to articulate any patterns, trends, relationships, or anomalies they observe, as well as any emerging questions or interpretations. Participants are informed that they may ask the facilitator for clarification if they encounter any uncertainty or difficulty using the tool.

    At the end of the 40 minutes, the experimenter initiates a short discussion with the participant to review and \am{together reformulate} the insights identified during the exploration time, \am{which will facilitate future coding}.

    \item \textbf{Post-Session Interview (15 min):} 
    Following the Exploratory Insight Generation phase, a semi-structured interview is conducted to gather participants’ reflections on their experience with ParcoursVis. The discussion focuses on their overall impression of the tool, the perceived value and relevance of the insights generated, and any difficulties encountered. 
    
    Participants are invited to review the insights they identified during the session and assess each one along three key dimensions: (1) \textit{Confidence}: how confident they are in the validity of the insight (2) \textit{Domain Value}: the perceived significance or usefulness of the insight within their professional context, rated on a scale from 1 (trivial observation) to 5 (hypothesis-generating insight); and (3) \textit{Actionability}: the extent to which the insight could lead to a concrete decision or change in practice. This reflective rating process supports both qualitative and quantitative understanding of insight utility.
    
    Additionally, participants are asked usability-focused questions, including which specific features of the visualization facilitated their understanding or helped uncover insights, and which aspects hindered their reasoning or caused confusion.

\end{itemize}

\subsection{Session 2: Remote Study}

\textbf{\am{Session 2.1}---Expert Review (Two Emergency Physicians):} 
In parallel, two senior emergency physicians not involved in the initial sessions serve as external reviewers. They independently evaluate the collected insights \am{(anonymized and with duplicates removed)} using the Insight characteristics (\autoref{sec:characteristics}). Their assessments serve as a reference to triangulate participant ratings of Session 2.2.

\noindent\textbf{\am{Session 2.2}---Participants' Review (Asynchronous):} 
After all participants have completed the initial session, they are invited to take part in a follow-up remote evaluation. In this session, each participant is presented with a curated set of insights gathered across all sessions (anonymized and \am{with duplicates removed}). They are asked to assess each insight along Insight characteristics (see~\autoref{sec:characteristics}). This cross-evaluation enables comparative validation across participants, helping to identify patterns of consensus and variability in perceived insight quality.

\section{Planned Analyses}

To comprehensively assess the \textit{utility} and \textit{analytical value} of insights generated through ParcoursVis, as well as the expert's strategy when generating observations, we combine both qualitative and quantitative measures. These include established insight characteristics and coding procedures inspired by prior work~\cite{6876022, taylor2023evaluating, saraiya2005insight}.

\subsection{Insight Characteristics}
\label{sec:characteristics}
\am{We recall that} each individual insight is defined as a discrete observation or unit of discovery made by a participant during their exploration of the visualization~\cite{saraiya2005insight}. The following characteristics are assessed by all participants during Session 2.1:

\begin{itemize}
    \item \textbf{Directed vs.\ Unexpected:} Insights are labeled as \textit{directed} if they confirm or disprove a prior expectation or hypothesis, or \textit{unexpected} if they emerge spontaneously during exploration.
    
    \item \textbf{Hypothesis Generation:} Whether the insight leads to a new hypothesis, direction of inquiry, or research question.
    
    \item \textbf{Correctness:} Whether the insight accurately reflects the data, as validated by expert reviewers.
    
    \item \textbf{Breadth vs.\ Depth:} \textit{Breadth} insights offer general patterns or overviews, while \textit{depth} insights involve focused, specific observations about attributes or relationships.
    
    \item \textbf{Domain Value:} Importance or relevance of the insight, rated on a 5-point Likert scale:
    \begin{enumerate}
        \item Trivial
        \item Low importance, marginally useful
        \item Moderate importance, potentially useful
        \item High importance, with clear relevance
        \item Critical; highly significant or hypothesis-generating
    \end{enumerate}
    
    \item \textbf{Actionability:} Degree to which the insight leads to or informs practical actions or decisions in a clinical setting (rated qualitatively and optionally on a Likert scale).
\end{itemize}

\subsection{Insight Coding and Categorization}

During Session 2.1, insights are qualitatively coded by two independent clinical experts using thematic coding methods~\cite{6876022}. \am{Coding is done on the collaboratively (participant and experimenter) reformulated set of insights produced in Session 1, during which redundant insights were discarded. Experts will review this curated list of insights and will focus on:}

\begin{itemize}[nosep]
    \item Clinical relevance; we aim to show where ParcoursVis can have a practical importance in real-world decision-making.

    \item Connections to system features to identify which parts of the system are most helpful or confusing.

\end{itemize}

In addition to predefined categories, the analysis remains open to the emergence of unanticipated themes through coding.
\am{Each observation will be classified into identified insight categories (e.g., clinically relevant, feature-linked). For example, a participant might note that a sequence is composed of repeated transfers between two ED zones, such as treatment and observation areas (e.g., the alternation of two differently colored nodes within a branch of the Icicle Tree in the Main View); this observation would be classified as clinically relevant and associated to the main visualization and/or the prefix-tree navigation features. Another possible case could be identifying a pattern of rapid discharges (i.e., patients being quickly sent back home) following specific triage scores (e.g., filtering on a specific score shows sequences of shorter durations in the Main View Icicle Tree); this would be classified as a clinically relevant observation tied to patient attribute filtering and/or the main visualization.}

\subsection{Mental Model}

We will analyze participants’ discourse to infer their underlying mental models and to determine whether they engage in top-down, hypothesis-driven reasoning or bottom-up, data-driven sensemaking processes~\cite{pirolli2005sensemaking,klein2007data}. \am{For example, a participant might begin exploration by hypothesizing that critical care areas have short stays (top-down) and then search for confirming or contradicting cases, whereas another participant could start by scanning the most frequent pathways without a specific hypothesis (bottom-up). Observing such differences in reasoning styles will inform our coding of cognitive strategies.} 

This analysis focuses on identifying markers such as explanatory or causal language, as well as indicators of opportunistic exploration and moments of sudden insight ``aha'' moments~\cite{klein2007data}. 
By aligning these cognitive strategies with user interaction patterns, we aim to refine the design of ParcoursVis to better accommodate users' reasoning styles, thereby enhancing overall usability. Furthermore, this analysis can contribute to defining the system’s appropriate scope of use and clarifying its position within the broader process of optimizing care pathways. It will also contribute to documenting the exploratory process for other health professionals.

\section{Pilot Results and Envisioned Study Refinements}

\am{The piloting of the study presented in~\autoref{sec:studyDesign} was conducted with the ED use case presented \autoref{sec:usecases}.  From the sixteen Parisian hospitals, we use data from three comparable hospitals (i.e., with a similar occupation for each zone, assuming an equivalent usage), covering 476,109 visits in about two years.}

\am{We interviewed two emergency physicians: one senior (+25 years of experience, head of an emergency service) and one junior (2 years of experience in ED). While their difference in seniority did not affect their ability to understand the data, the senior physician was more reflexive in comparing patterns across hospitals, drawing on her experience working in several institutions within the Parisian area.}
Unlike the earlier clinical collaborators who helped refine ParcoursVis and were familiar with data availability constraints, these participants had limited knowledge of the underlying data workflow. This difference in familiarity highlighted several key lessons that are informing adjustments to our study design.
\am{In this pilot, we conducted only the first session of our planned multi-session protocol. The preliminary insights identified during this session were not shared back with participants, as they were limited in number and associated with needs and opportunities listed below. Given the time constraints of our participants, we chose to refine the protocole and the system before proceeding further.}

\begin{enumerate}
\item \textbf{The need for richer contextual data:}
Both participants expressed difficulty forming actionable insights due to missing contextual attributes in the visualization. For example, the ``Waiting Area'' event was perceived as ambiguous: patients could be waiting for a variety of procedures or consultations, which holds important clinical implications. This feedback highlights a broader issue of limited contextual data availability in EHR systems, as noted in the EHR STAR report~\cite{Wang:2022:CGF}. Additionally, participants emphasized the importance of knowing the admission path (i.e., the pathway through which patients enter the ED), as well as the discharge outcome, such as whether the patient was sent home or admitted to another hospital unit. They also raised questions about in-pathway mortality, which should be included as a filtering attribute for specific explorations. 
This kind of feedback illustrates a \textit{virtuous data circle}, where exploratory interaction with the tool not only generates clinical insights but also helps uncover gaps in the available data. In other words, this process supports the identification of contextual variables that could enhance subsequent analysis. Fortunately, we expect to gain access to this additional data in the coming months. Future iterations of the tool will incorporate these elements to enrich event semantics and better support meaningful clinical insights.

\item\textbf{Separating training from exploratory analysis:}
Participants were asked to perform an exploratory analysis immediately after being introduced to the tool, which proved to be too abrupt. We observed that familiarity with the interface and with the data needed more time to develop. This suggests the study would benefit from being split into two sessions: an initial training and usability session, followed days or weeks later by a session focused on open-ended exploration. This redesign aims to better simulate real-world adoption patterns and allow participants to engage more meaningfully with the data. The MILCS longitudinal methodology also advised this multi-stage approach~\cite{MILCS}.

\item\textbf{Addressing hospital-specific data specification:}
Finally, in-field observation revealed the strong influence of local hospital practices on how ED zones are used and labeled. While the information system is standardized, each hospital adapts it differently, introducing inconsistencies in zone semantics. To address this, we will conduct site visits to map zone usage more precisely. In the short term, this requires additional engineering effort to ensure valid data specification.
\am{More broadly, this highlights the need for future versions of ParcoursVis to support domain experts in defining or adjusting event types during the data preprocessing stage, with underlying engineering work both in the data processing pipeline and tool documentation to ensure that zone semantics are accurate for each hospital and easily adjustable based on on-site observations. Data are not given; they should be collected and constructed with caution to build effective support tools.}
\end{enumerate}

\noindent These lessons collectively inform both the short-term refinements of our study protocol and the long-term design of the system. By adapting to the realities of clinical data and user workflows, we aim to support more accurate, meaningful, and context-aware exploratory analysis in real-world healthcare settings.

\section{Conclusion}

As part of our ongoing work on evaluating ParcoursVis, we present an approach for assessing mental models and the insights generated by expert users when exploring EHR data. 
We have so far piloted this approach with two expert healthcare professionals to refine both the system and the evaluation protocol.
This process revealed data needs, i.e., contextual attributes for interpreting care pathways such as patient admission routes, discharge specificities, and in-pathway mortality. These findings reinforce the importance of data consistency in supporting actionable insights and inform our request for expanded access and collection of key EHR variables.

We formalized an evaluation protocol designed to align future studies and bring visibility to the value of exploratory visualization tools.
\am{Although our study is grounded in the context of large-scale EHR exploration, the proposed methodology, i.e., combining insight-based evaluation with the analysis of cognitive strategies, is general and could be applied to other medical visualization systems aimed at open-ended, expert-driven analysis such as proton therapy planning~\cite{MUSLEH2023166} or other EHR visualization systems~\cite{Wang:2022:CGF}.}

We are currently working to incorporate the newly identified data into the system to enable more advanced, context-rich exploration. Our goal is to examine not only how insights are generated but also how they translate into hypotheses and contribute to the adoption of more effective clinical practices in real-world settings.

\section*{Acknowledgements}
We thank our reviewers for their valuable insights that helped improve this paper. This work was supported in part by a grant from the Health Data-Hub, and from the \href{https://www.inria.fr/en/urge}{URGE AP-HP/Inria project}.


\bibliographystyle{abbrv-doi-hyperref}

\bibliography{template}
\typeout{get arXiv to do 4 passes: Label(s) may have changed. Rerun}

\end{document}